\documentstyle[preprint,aps,prb]{revtex}

\begin{document}
\draft

\preprint{Submitted to \prb}

\title{Magnetoresistance of composite fermions at $\nu=1/2$}

\author{L.~P.~Rokhinson%
\footnote{Present address: Department of Electrical Engineering,
Princeton University, Princeton, NJ 08544} and V.~J.~Goldman}

\address{Department of Physics, State University of New York, Stony
Brook, NY 11974}

\date{\today}

\maketitle

\begin{abstract}

We have studied temperature dependence of both diagonal and
Hall resistivity in the vicinity of $\nu=1/2$.  Magnetoresistance was
found to be positive and almost independent of temperature:
temperature enters resistivity as a logarithmic correction. At the same
time, no measurable corrections to the Hall resistivity has been
found.  Neither of these results can be explained within the mean-field
theory of composite fermions by an analogy with conventional
low-field interaction theory.  There is an indication that interactions
of composite fermions with fluctuations of the gauge field may
reconcile the theory and experiment.

\end{abstract}

\pacs{PACS numbers: 73.40.Hm}

Experimentally, it has been known for some time that in
low disorder two-dimensional electron systems (2DES) at filling
factor $\nu=1/2$ the diagonal resistivity $\rho_{xx}$ remains finite at
low temperatures and exhibits a shallow minimum, while the Hall
resistivity $\rho_{xy}$ is nearly linear in magnetic field and does not
form a plateau.   An understanding of the phenomenon came with the
theory\cite{jain89,halperin93} of composite fermions (CFs), where
weakly interacting new particles -- composite fermions -- were
proposed\cite{halperin93,kalmeyer-kivelson} to form a metallic Fermi
liquid-like state near $\nu=1/2$.  In the mean-field approximation CFs
experience a reduced effective magnetic field $B_{cf}=B-2n\phi_{0}$,
where $n$ is the electron (and CF) concentration, and $\phi_{0}=h/e$
is the flux quantum. At $\nu=1/2$ the external magnetic field is fully
cancelled and $B_{cf}=0$; it has been shown experimentally
\cite{goldman-willett-kang} that some properties of a Fermi liquid are
preserved for CFs, in particular, a reasonably well defined Fermi
surface.

Despite some similarity between
$\nu=1/2$ and $B=0$ phenomenology, there are apparent differences in
transport properties.  For example, magnetoresistance is negative near
$B=0$, while it is positive near $\nu=1/2$.
Magnetoresistance at low $B$ has been a
powerful tool in the study of weak localization and electron interaction
effects.  This method relies on the prediction of the classical Drude
model that $\rho_{xx}$ is not affected by magnetic field,  while
$\sigma_{xx} = \rho_{xx}/(\rho_{xx}^2 + \rho_{xy}^2)$ displays negative
magnetoconductance via $\rho_{xy}\propto B$.  Any magnetoresistance
then results from quantum corrections to the conductivity tensor,
which, in general, have different $B$- and $T$-dependence than Drude
$\sigma_{xx}^0$ and $\sigma_{xy}^0$ and, thus, can be separated.
Altshuler--Aronov quantum correction to conductivity
$\Delta\sigma_{xx}^{AA}$ due to interaction effects has
a logarithmic temperature dependence\cite{altshuler80} and is field
independent at low $B$ because the correction
to Hall conductivity $\Delta\sigma_{xy}^{AA}=0$
\cite{houghton-girvin}.  Neglecting the weak localization
contribution, for electrons at low $B$ the resulting quantum
magnetoresistance $\Delta\rho_q = \rho_{xx}(B)-
\rho_{xx}(0)\approx\rho_{xy}^2\Delta\sigma_{xx}^{AA}$ (for
$\Delta\sigma_{xx}^{AA}\ll\sigma_{xx}^0$) is negative, because
$\Delta\sigma_{xx}^{AA}<0$\cite{houghton-girvin,paalanen83}.

We have reported recently observation of a logarithmic
correction to  the conductivity of CFs $\sigma_{xx}^{cf}$ at
$\nu=1/2$ and attributed it to the short-range interaction between CFs
\cite{rokhinson95}.  An enhancement of the coupling constant,
compared to the low-field regime, was found recently to be a result of
an interaction between CFs via the gauge field fluctuations
\cite{khveshchenko96}. Naively, one may also expect that this effect
should lead to a negative magnetoresistance, in analogy to the low-$B$
case.  However, experimentally positive magnetoresistance and no
correction to the Hall resistivity are measured near $\nu=1/2$.
Thus, non--zero correction $\Delta\sigma_{xy}^{cf}\neq0$, in addition
to $\Delta\sigma_{xx}^{cf}\neq0$, both $B$-dependent, is required to
reconcile measured corrections to $\rho_{xx}$ and $\rho_{xy}$ with the
constrains imposed by the matrix inversion of transport coefficients.

We have studied several samples fabricated from high mobility
($\mu\approx2\times10^6\ \text{cm}^2/\text{V s}$)
$\text{GaAs/Al}_{x}\text{Ga}_{1-x}\text{As}$ heterojunction wafers.
The wafers have double Si $\delta$-doping, the first layer is separated
from the 2DES by a $d_s=120$ nm thick spacer. 2DES with densities
0.4 and 1.2$\times$10$^{11}$ cm$^{-2}$ were prepared by illuminating a
sample with red light.  The temperature was measured with a
calibrated Ruthenium Oxide chip resistor. Measurements were done
in a top-loading into a mixture dilution refrigerator using a
standard lock-in technique. Samples were patterned in either Corbino or
Hall bar geometry.

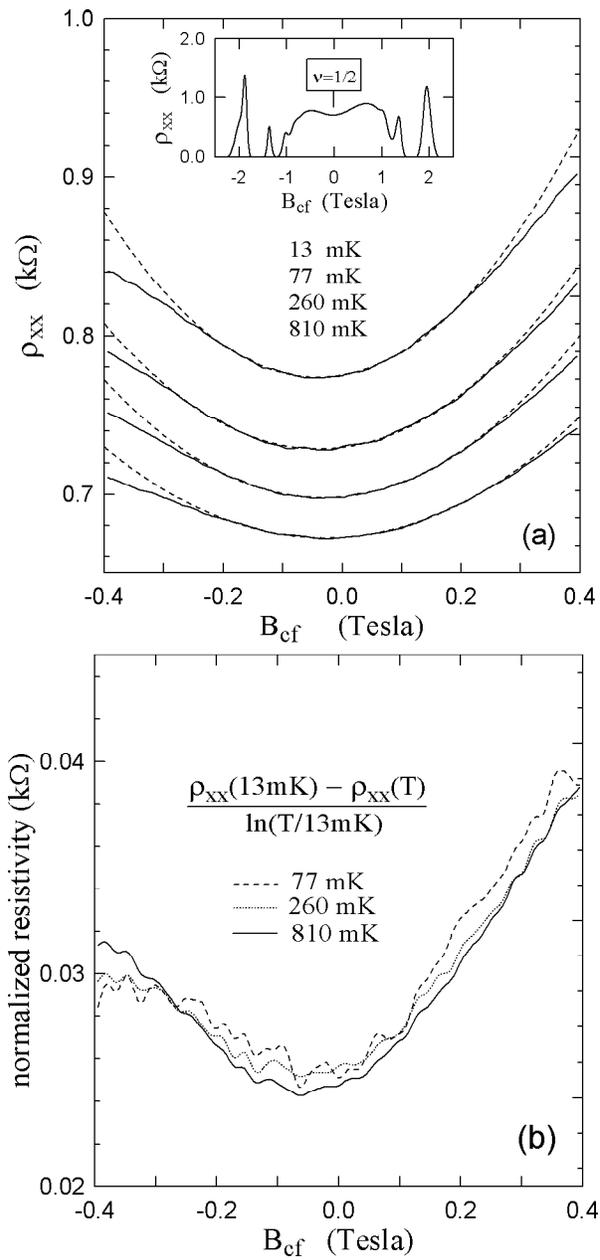
\begin{figure}
\label{rhoxx}
\caption{a) Magnetoresistivity data $\rho_{xx}$ vs $B_{cf}$ near
$\nu=1/2$ for $T=13$, 77, 260 and 810 mK (from top to bottom). Dashed
lines are polynomial fits Eq.~(\protect\ref{polynom}) in the range
$|B_{cf}|<0.12$ Tesla.  Resistivity in a larger field range is
shown in the inset. b) The scaling of the difference between $\rho_{xx}$
at 13 mK and other temperatures, normalized by the log of the ratio of
temperatures.}
\end{figure}

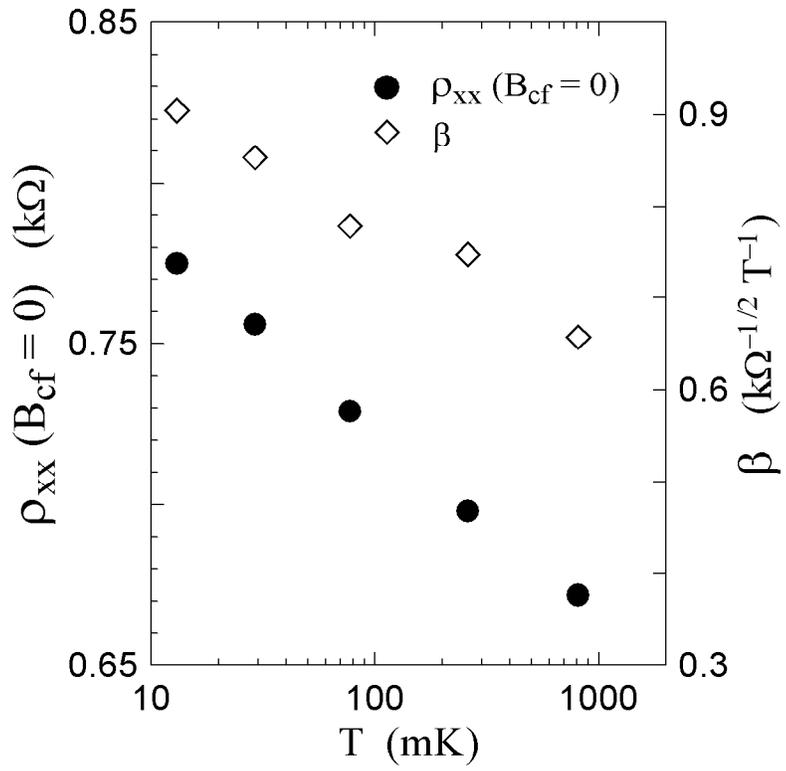
\begin{figure}
\label{rhobeta}
\caption{The $\nu=1/2$ resistivity $\rho_{xx}(0,T)$
plotted as a function of temperature.  The coefficient $\beta$,
defined in Eq.~(\protect\ref{polynom}), is obtained from the fits in
Fig.~\protect\ref{rhoxx}.}
\end{figure}

Representative magnetoresistivity data $\rho_{xx}(B_{cf},T)$ near
$\nu=1/2$ are plotted in Fig.~\ref{rhoxx}(a) (note that $\rho_{xx}^{cf}=
\rho_{xx}$).  Magnetoresistance is positive near $\nu=1/2$ and
depends on temperature weakly. A remarkable result is that $\rho_{xx}$
at a given $B_{cf}$ changes logarithmically with temperature for
13 mK $<T<$ 1000 mK. A simple function
\[
\left[ \rho_{xx}(B_{cf},T) - \rho_{xx}(B_{cf},T_1) \right]
/ \ln(T_1/T)
\]
collapses $\rho_{xx}$ vs $B_{cf}$ traces at different temperatures $T$
into a single curve [Fig.~\ref{rhoxx}(b)].  Such a scaling requires
that both, the $B_{cf}=0$ part of resistivity $\rho_{xx}(0,T)$,
and the part responsible for the magnetoresistance, have terms
proportional to $\log T$.
We fit the data with a polynomial
\begin{equation}
\rho_{xx}(B_{cf},T)=
\rho_{xx}(0,T) + \alpha(T)B_{cf} + \beta(T)B_{cf}^2
\label{polynom}
\end{equation}
(dashed lines in Fig.~\ref{rhoxx}a) in a
classically weak-field region for CFs $\rho_{xy}^{cf}\leq\rho_{xx}^{cf}$
(corresponds to $|B_{cf}|\leq0.12$ T for the sample in
Fig.~\ref{rhoxx}).  Value of $\alpha\neq0$ corresponds to a known term
in $\rho_{xy}$ proportional to $B\frac{d\rho_{xy}}{dB}$
\cite{chang-stormer}.  As it is expected from the above analysis, both
$\rho_{xx}(0,T)$ and $\beta(T)$ change logarithmically with
temperature (Fig.~\ref{rhobeta}). Zero--field CF
conductivity $\sigma_{xx}^{cf}(0,T) = 1/\rho_{xx}(0,T)$ has a
negative logarithmic $T$-dependent correction, which has been
attributed to interaction effects between CFs, analogous to the
Altshuler-Aronov type localization correction for electrons at low
magnetic field\cite{halperin93,rokhinson95}.  However, as apparent from
Fig.~\ref{rhoxx}, there is a positive magnetoresistance near
$B_{cf}=0$, in a stark contrast to the negative magnetoresistance near
$B=0$.

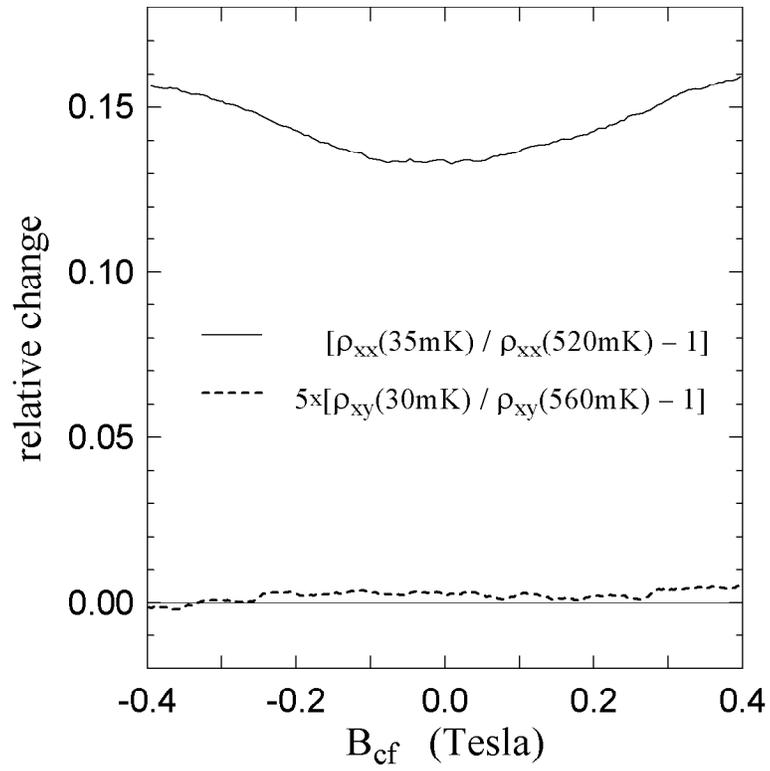
\begin{figure}
\label{relrho}
\caption{Relative change of $\rho_{xx}$ and $\rho_{xy}$ with
temperature.  Note that the change in $\rho_{xy}$ is multiplied by a
factor of 5.}
\end{figure}

In contrast to the low-field regime, we have found no deviation of
Hall resistivity $\rho_{xy}$ from its free-electron value
$\rho_{xy}^0=B/en$ near $\nu=1/2$ (electron concentration $n$ is
determined from Shubnikov-de Haas oscillations with
2\% accuracy). A direct comparison of $\rho_{xy}$ at $35$ and  560 mK
 shows (Fig.~\ref{relrho}) that there is no $T$-dependent correction to
$\rho_{xy}$ within experimental error of 0.1\% in
the range $|\omega_c^{cf}\tau|<3$. This value should be contrasted with
$\approx15\%$ change of $\rho_{xx}$. Thus, we conclude that
$\Delta\rho_{xy}=0$ near $\nu=1/2$.

Within the mean field theory transport properties of 2DES  near
$\nu=1/2$ closely resemble those near $B=0$.  Let us examine mechanisms
which may lead to the positive magnetoresistance within the
$\{\nu=1/2\}\leftrightarrow \{B=0\}$ analogy.  At low $B$, there are no
corrections to $\rho_{xy}$ due to weak localization \cite{lee85}.  Near
$\nu=1/2$, the disorder--induced fluctuations of electron density
$\delta n$ produce static fluctuations of the gauge field $\delta
B_{cf}=2\delta n\phi_0$, and the first--order correction to $\rho_{xx}$
is suppressed\cite{kalmeyer92}.  The second--order correction is
$\sim100$ times less than the measured logarithmic term in
$\rho_{xx}(0,T)$ \cite{aronov94}.  Also, static fluctuations of
the gauge field would suppress quantum interference at
$\omega_c^{cf}\tau\approx 1$, although the positive magnetoresistance
is observed to much higher effective magnetic fields.

Another possible source for positive magnetoresistance is a
classical correction to the Drude resistivity $\rho_{xx}^{0}$, which
results from the fact that an average size of potential fluctuations is
larger than the Fermi wavelength.  Simple arguments \cite{drude-note}
lead to the following positive quadratic in $B_{cf}$ correction to
$\rho_{xx}^{0}$:
\begin{equation}
\Delta\rho_{cl}\propto
\rho_{xx}^0\left(\frac{d_s}{r_c}\right)^2,
\label{rho_cl}
\end{equation}
where $d_s$ is the spacer thickness and $r_c={{\hbar k_F}\over{e
B_{cf}}}$ is the cyclotron radius.  Recent experiments\cite{mancoff95}
show that, in the presence of a spatially non-uniform magnetic field, a
positive magnetoresistance is observed in 2DES at low magnetic
fields.  However, the classical
magnetoresistance has been calculated for $T=0$ and thus does not
have any temperature dependence.  We do not expect appreciable
temperature dependence for this scattering mechanism, at least for
$T<0.5$ K, when phonon scattering is negligible\cite{kang95},
inconsistent with the observed $\log T$ dependence of resistivity.
Thus, the classical correction alone cannot explain the experimental
results.

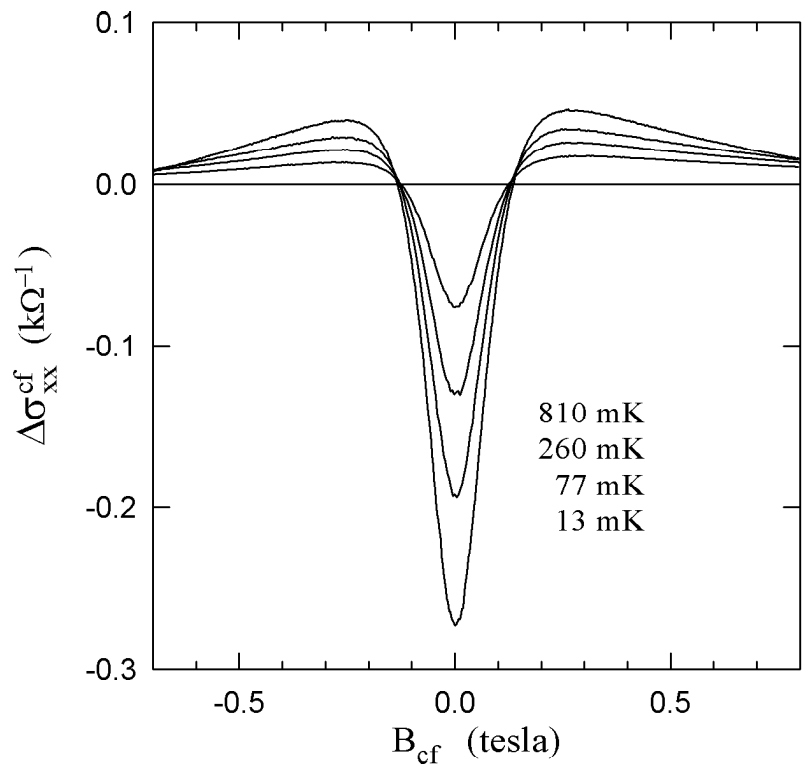
\begin{figure}
\label{deltasigma}
\caption{Deviation of $\sigma_{xx}^{cf}$ from the Drude value is shown
near $\nu=1/2$ for $T=13$, 77, 260 and 810 mK. Note the change of sign
of $\Delta\sigma_{xx}^{cf}$ at $\rho_{xx}^{cf}=\rho_{xy}^{cf}$.}
\end{figure}

The logarithmic temperature dependence of $\beta(T)$ strongly suggests
that the positive quadratic magnetoresistance originates from the
interaction effects between CF's. This conclusion is further supported
by the observation that both $\rho_{xx}(0,T)$ and $\beta(T)$ deviate
from $\log T$ dependence at about the same $T$.
However, matrix inversion  of transport
coefficients, combined with Onsager relations and experimental
observations that (i) $\Delta\rho_{xy}/\rho^0_{xy}\ll
\Delta\rho_{xx}/\rho_{xx}^0$ (Fig.~\ref{relrho}), and (ii)
both $\rho_{xx}$ and $\rho_{xy}$ are non-singular near
$\nu=1/2$, impose certain constrains on the corrections to the Drude
conductivity tensor.  Assuming that both corrections are small
($\Delta\sigma_{xx}\ll\sigma_{xx}$ and
$\Delta\sigma_{xy}\ll\sigma_{xy}$) they can be expressed in the
following form:
\begin{mathletters}
\begin{eqnarray}
\Delta\sigma_{xx}^{cf}(B_{cf},T)&\approx&
  f(\gamma)(1-\gamma^2)\Delta\sigma_{xx}^{cf}(0,T)\\
\Delta\sigma_{xy}^{cf}(B_{cf},T)&\approx&
  2\gamma f(\gamma)\Delta\sigma_{xx}^{cf}(0,T),
\end{eqnarray}
\label{ds}
\end{mathletters}
where $\gamma=\rho_{xy}^{cf}/\rho_{xx}^0 \propto B_{cf}$
($\gamma=\omega_c^{cf}\tau$ in the Drude model), $f(\gamma)$ is an
even smooth function of $B_{cf}$ and $f(0)=1$.  Note, that the $B-$ and
$T-$dependencies are separated, and $T$ enters only through
the zero-field correction to diagonal conductivity
$\Delta\sigma_{xx}(0,T)$.  Indeed, experimentally determined
$\Delta\sigma_{xx}$ and $\Delta\sigma_{xy}$ are both $B$-dependent and
$\Delta\sigma_{xx}$ changes sign at $\rho_{xx}\approx\rho_{xy}^{cf}$
(Fig.~\ref{deltasigma}).

All these findings contradict the results
of the conventional low-field interaction theory, which predicts
$\Delta\sigma_{xy}=0$ and a field independent $\Delta\sigma_{xx}$
\cite{lee85}.  A recent theory investigated interaction
effects between CFs in the presence of disorder beyond the mean-field
approximation\cite{khveshchenko96a}.  The calculated corrections are in
agreement with the above qualitative analysis [Eqs.~(\ref{ds})] with
$f(\gamma)\equiv 1$ and $\gamma\equiv \omega^{cf}_c \tau$.  These
corrections to conductivity lead to the  following corrections to
the resistivity tensor:
\begin{mathletters}
\begin{eqnarray}
\Delta\rho_{xx}(B_{cf},T)&\approx&
\Delta\rho_{xx}(0,T) [1+(\omega_c^{cf}\tau)^2]^2\\
\Delta\rho_{xy}(B_{cf},T)&\approx& - \rho_{xy}^0
[\Delta\rho_{xx}(0,T)/\rho_{xx}^0]^2
[1+(\omega_c^{cf}\tau)^2]
\label{drxy}
\end{eqnarray}
\label{dr}
\end{mathletters}
Qualitatively, Eqs.~(\ref{dr}) predict a positive
magnetoresistance and a vanishing term linear in
$\Delta\rho_{xy}$. However, thus calculated $\Delta\rho_{xx}$
overestimates $\beta$ from Eq.~(\ref{polynom}) by a factor of 20, if we
use $\omega_c^{cf}\tau=\rho_{xy}^{cf}/\rho_{xx}^0$, with
$\rho_{xx}^0=0.65$ k$\Omega$.  Also, a large quadratic correction
to the Hall resistivity, $\Delta\rho_{xy}/\rho_{xy}^0>2.5\%$, estimated
from Eq.~(\ref{drxy}), is inconsistent with experiment ($<0.1\%$, see
Fig.~\ref{relrho}).

Our main results can be summarized as follows: (i)
experimentally, the resistivity has a logarithmic  temperature
dependence near $\nu=1/2$, which implies that both $B$-independent
resistivity and magnetoresistance have $\log T$ dependence, and (ii)
there is no measurable correction to the classical Hall resistivity
near $\nu=1/2$.  From analysis of possible mechanisms which may lead to
a positive magnetoresistance, we conclude that the observed
$T$-dependencies cannot be explained within the mean-field theory of
CFs.  The similar $\log T$ dependence of resistivity at $\nu=1/2$ and
of magnetoresistance suggests that both corrections have the same
physical origin, namely, interactions between CFs.

We are grateful to M. Shayegan for MBE material and to I. Aleiner for
stimulating discussions.  This work was supported in part
by NSF under Grant  No. DMR-9318749.

\end{document}